\documentstyle[12pt,psfig]{article}

\def\Journal#1#2#3#4{{#1} {\bf #2}, #3 (#4)}

\def\PLB{{\em Phys. Lett.}  B}
\def\PRL{\em Phys. Rev. Lett.}
\def\PRD{{\em Phys. Rev.} D}

\def\be{\begin{equation}}
\def\ee{\end{equation}}
\def\bea{\begin{eqnarray}}
\def\eea{\end{eqnarray}}
\newcommand{\lsim}{\mathrel{\lower4pt\hbox{$\sim$}}
\hskip-12.5pt\raise1.6pt\hbox{$<$}\;}

\newcommand{\gsim}{\mathrel{\lower4pt\hbox{$\sim$}}
\hskip-12.5pt\raise1.6pt\hbox{$>$}\;}

\def\qwe{\zeta}

\begin{document}
\noindent \hspace*{10cm}UCRHEP-T190\\
\noindent \hspace*{10cm}May 1997\\

\begin{center}
{\bf FLAVOR CHANGING TRANSITIONS AND THE TOP QUARK IN A TWO HIGGS DOUBLETS MODEL; APPLICATION TO $e^+e^-\to t\bar c\nu_e\bar\nu_e$, $t\bar
ce^+e^-$, $t \bar c Z$, $t\to cW^+W^-$ AND $t\to cZZ$}   \footnote{Talk presented at the first FCNC symposium, February 19-21, 1997, Santa Monica, CA USA. Work done in collaboration with G. Eilam, A. Soni and J. Wudka.}
\vspace{.7in}

S. Bar-Shalom \\
Department of Physics, University of California, \\
Riverside CA 92521.
\end{center}
\vspace{.35in}

\begin{center}
{\bf Abstract}\\
\end{center}

Extension of the SM with one extra Higgs doublet may give rise to flavor-changing-scalar coupling of a neutral Higgs to a pair of top-charm quarks. This coupling gives rise to a large tree-level effective $W^+W^-;ZZ-{\rm Higgs}-t \bar c$ interaction. We find that the reactions $e^+e^-\to t\bar c\nu_e\bar\nu_e$, $t\bar
ce^+e^-$, $t \bar c Z$ and the two rare top decays $t\to cW^+W^-$, $t\to cZZ$ are very sensitive probes of such an effective interaction. The most promising ones, $e^+e^-\to t\bar c\nu_e\bar\nu_e$, $t\bar
ce^+e^-$, may yield several hundreds and up to thousands of such 
events at the Next Linear Collider with a center of mass energy of
$\sqrt{s}=0.5$--2 TeV if the mass of the light neutral Higgs is a few hundred GeV\null. We also find that the rare decays $t\to cW^+W^-$ and $t\to cZZ$ may be accessible in the LHC if the mass of the light neutral Higgs lies in the narrow window $150~{\rm GeV} \lsim m_h \lsim 200~{\rm GeV}$. 

\pagebreak  
\section{Introductory Remarks}

The reactions $\sigma^{Ztc} \equiv \sigma(e^+e^- \to t \bar c Z + \bar t c Z)$ and in particular 
$\sigma^{\nu \nu tc} \equiv \sigma(e^+e^- \to t \bar c \nu_e {\bar \nu_e}+ \bar t c \nu_e {\bar \nu_e})~,~ 
\sigma^{eetc} \equiv \sigma(e^+e^- \to t \bar c e^+ e^- + \bar t c e^+ e^-)$, 
via $W^+ W^-$ or $ZZ$ fusion, respectively, are extremely sensitive probes for investigations of FC currents at the Next Linear Collider (NLC). Also, in a very narrow window of the Higgs mass detection of the rare top decays $t\to cW^+W^-$ and $t\to cZZ$ may be possible at the Large Hadron Collider.\cite{hepph9703221} 

A two Higgs doublet model with flavor-changing-scalar-transitions (FCST),
often called Model~III,\cite{model3} can give rise to large FCST in top quark systems.\cite{model3phen} In this model there are two neutral scalars, denoted here by $h$ and $H$, that can couple to vector bosons.  
Within Model~III the above reactions can occur at the tree-level via the neutral scalar exchanges in Fig.~1. In the Cheng-Sher Ansatz (CSA),\cite{csa} the couplings of the neutral scalars to fermions are given by
$\xi_{ij}^{U,D}=g_W \left({\sqrt {m_im_j}}/m_W \right) \lambda_{ij}$
and the relevant terms of the Model~III Lagrangian can then be written as:
\begin{eqnarray}
{\cal L}_{{\cal H}tc}&=&-\frac{g_W}{\sqrt 2} \frac{{\sqrt {m_t m_c}}}{m_W} f_{\cal H} {\cal H} \bar t (\lambda_R+i\lambda_I \gamma_5) c  \label{htc}~,\\
{\cal L}_{{\cal H} VV}&=&-g_W m_W C_V c_{\cal H} {\cal H} g_{\mu\nu} V^{\mu} V^{\nu} \label{hww}~,
\end{eqnarray}
\noindent where ${\cal H}=h~{\rm or}~H$ and for simplicity we choose $\lambda_{tc}=\lambda_{ct}=\lambda$.\footnote{Existing experimental information does not provide
any useful constraints on $\lambda_{tc}$; in particular, we may well have
$\lambda_{tc}\sim {\cal O}(1)$.} We furthermore break $\lambda$ into its real and imaginary parts, $\lambda=\lambda_R+i\lambda_I$. Also in Eqs.~\ref{htc} and \ref{hww}    
$C_{W;Z}\equiv1;m_Z^2/m_W^2 \label{cwz}$, 
$f_{h;H}\equiv \cos{\tilde {\alpha}};\sin{\tilde {\alpha}} \label{fhh}$ and
$c_{h;H} \equiv \sin{\tilde {\alpha}};-\cos{\tilde {\alpha}} \label{chh}$ where the mixing angle $\tilde {\alpha}$ is determined by the Higgs potential.
  
\begin{figure}
%\rule{5cm}{0.2mm}\hfill\rule{5cm}{0.2mm}
%\vskip 2.5cm
%\rule{5cm}{0.2mm}\hfill\rule{5cm}{0.2mm}
\psfig{figure=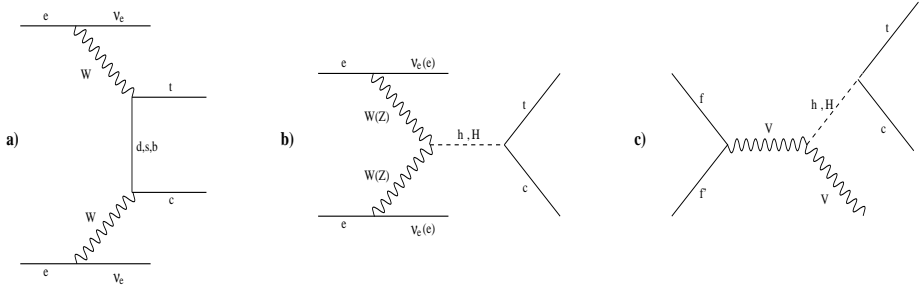,height=1.5in,width=4.7in}
\caption{(a) The Standard Model diagram for $e^+e^- \to t \bar c \nu_e {\bar {\nu}}_e$; (b) Diagrams for $e^+e^- \to t \bar c \nu_e {\bar {\nu}}_e (e^+e^-)$ in Model~III; (c) Diagram for $f \bar {f'} \to t \bar c V$ in Model~III.
\label{fig1}}
\end{figure}

\section{Production of $t \bar c$ Pairs Through Vector Boson Fusion}

In Model~III, $VV \to t \bar c$ ($V=W~or~Z$) proceeds at tree-level 
via the s-channel neutral Higgs exchange of diagram b in Fig.~1.           
Neglecting the SM diagram,\footnote{It is interesting to note that in the 
SM, the parton level reaction $W^+W^- \to t \bar c$ can proceed at tree-level, 
via diagram a in Fig.~1. However, numerically, $ \hat \sigma_{\rm SM} $ is found to be too small to be of experimental
relevance as it suffers from a severe CKM suppression.} the parton-level cross-section 
${\hat \sigma}_V \equiv {\hat \sigma}(V^1_{h_{V^1}}V^2_{h_{V^2}} \to t \bar c)$
is given by:
\begin{eqnarray}
{\hat \sigma}_V &= 
\frac{\left( \sin 2{\tilde \alpha } \right)^2 N_c \pi \alpha^2}{4 \hat s
\beta_V s^4_W}  \left(\frac{m_V}{m_W}\right)^4 
| \epsilon^{V^1}_{h_{V^1}} \cdot \epsilon^{V^2}_{h_{V^2}}|^2  
|\Pi_h - \Pi_H |^2 \times 
\nonumber \\
&  \sqrt { \Delta_t \Delta_c a_+ a_- } ( a_+ \lambda_R^2 + a_- \lambda_I^2 )    \label{vvhtc}~,
\end{eqnarray}
\noindent where $ a_\pm = 1 - ( \sqrt {\Delta_t} \pm \sqrt{\Delta_c} )^2$, 
$\beta_\ell \equiv \sqrt {1-4 \Delta_\ell^2}$ and $\Delta_{\ell} \equiv m_{\ell}^2/\hat s$. Also:
\begin{eqnarray}
\Pi_{\cal H} = \left(1 -\Delta_{\cal H} +i \sqrt {\Delta_{\cal H} \Delta_{\Gamma_{\cal H}}} \right)^{-1} ~~,~~ \Delta_{\Gamma_{\cal H}} \equiv \Gamma_{\cal H}^2 {\hat s}^{-1} \label{pih}~.
\end{eqnarray}
\noindent The corresponding full cross sections $\sigma^{\nu \nu tc}$ and $\sigma^{eetc}$ can then be calculated by folding in the distribution
functions $f^V_{h_V}$, for a vector boson $V$ with 
helicity $h_V$. 

Figure 2 shows the dependence of the scaled cross-section (SCS)
$\sigma^{\nu \nu tc}/\lambda^2$ on the mass of the light Higgs $m_h$ for 
four values of $s$ and for $m_H=1$ TeV, $\tilde {\alpha}=\pi/4$. The cross-section peaks at 
$m_h \simeq 250$ GeV and drops as the mass of the light Higgs approaches 
that of the heavy Higgs. Indeed as $m_h \to m_H$, 
$\sigma^{\nu \nu tc}/\lambda^2 \to 0$ as expected when $\tilde {\alpha} =\pi/4$ for which the couplings $htc$ and $Htc$ are identical. However, this ``GIM like'' cancelation does not operate when $\tilde {\alpha} \neq \pi/4$ for which $\sigma^{\nu \nu tc}/ \lambda^2$ can stay at the fb level even 
for $m_h \to m_H$.\cite{hepph9703221} The $ZZ$ fusion cross-section, $\sigma^{eetc}$, exhibits the same behavior though suppressed by about one order of magnitude.  

\begin{figure}
\psfig{figure=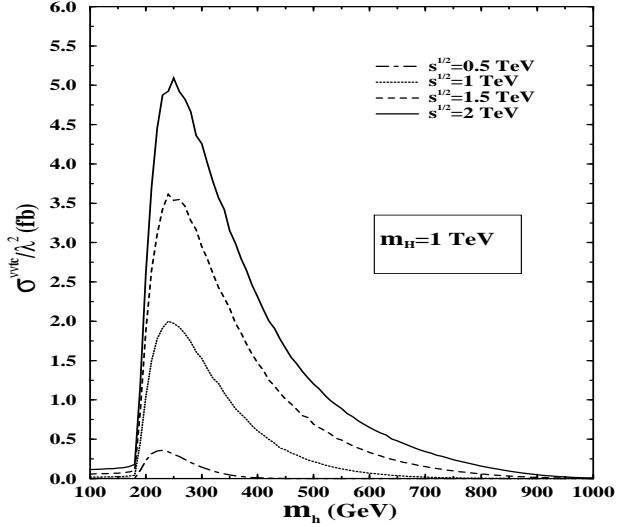,height=2.0in,width=4.0in}
\caption{The cross-section $\sigma (e^+e^- \to t \bar c \nu_e {\bar {\nu}}_e +  \bar t c \nu_e {\bar {\nu}}_e)$
in units of $\lambda^2$ as a function of $m_h$ 
for center of mass energies of 0.5,1,1.5 and 2 TeV. 
\label{fig2}}
\end{figure}

\section{Production of $Zt \bar c$ via $e^+e^- \to Z t \bar c$}

Within Model~III, the reaction $f \bar {f'} \to V t \bar c$ ($V=Z,W^+$ or $W^-$ depending on the quantum numbers of $f \bar {f'}$) proceeds at tree-level via the Feynman diagram c in Fig.~1. Of course, disregarding the incoming $f \bar {f'}$ fermions, this reaction is directly related to the sub-process $VV \to {\cal H} \to t \bar c$. We can therefore express the cross-section $\sigma ( f \bar {f'} \to V t \bar c)$ in terms of the hard cross-section ${\hat {\sigma}}_V$ given in Eq.~\ref{vvhtc}: 
\begin{eqnarray}
\sigma ( f \bar {f'} \to V t \bar c) && = \frac{\alpha}{6 \pi (\sin 2 \theta_W)^2} \Delta_V \Pi_V^2 \left( (a_L^{f(V)})^2+ (a_R^{f(V)})^2 \right) \times \nonumber\\
&& \int_1^{(\Delta_t^{-1/2} - \qwe_V)^2} dz \omega_1 \omega_2 \frac{\omega_1^2 +12 \Delta_V}{\omega_2^2 + 12 \qwe_V^4} ~ \sum_{h_{V^1},h_{V^2}} 
\left. {\hat {\sigma}}_V \right|_{\hat s = m_t^2 z} \label{eeztc}~. 
\end{eqnarray}
\noindent Here $\Delta_{\ell} \equiv m_{\ell}^2/s$ ($s$ being the center of mass (c.m.) energy of the colliding $f \bar {f'}$ fermions) and $\Pi_V=(1- \Delta_V)^{-1}$. Also, $\qwe_{\ell} \equiv m_{\ell}/m_t$ and $\omega_1$, $\omega_2$ are function of $z$ given by: 
$\omega_1= \sqrt { \left(1-(\sqrt {\Delta_V} + \sqrt {\Delta_t z}) \right) 
\left( 1-(\sqrt {\Delta_V} - \sqrt {\Delta_t z}) \right)}$, 
$\omega_2 = z \sqrt {1- 4 z^{-1} \qwe_V^2}$.

We wish to concentrate on the cross-section $\sigma^{Ztc} \equiv \sigma(e^+ e^- \to Zt \bar c + Z \bar t c)$ for which $V=Z$, $f=e^-$, $\bar {f'}=e^+$ and $a_L^{e(Z)}=1/2 - s_W^2$, $a_R^{e(Z)}=s_W^2$. 
In Fig.~3 we present the SCS $\sigma^{Ztc}/ \lambda^2$ as a function of the light Higgs mass, $m_h$, for $\sqrt s=0.5,1,1.5$ and 2 TeV where $m_H=1$ TeV and $\tilde {\alpha}=\pi /4$.
We see that there is a significant difference between $\sigma^{\nu \nu tc}$ (and/or $\sigma^{eetc}$) and $\sigma^{Ztc}$; namely, while $\sigma^{\nu \nu tc}$ grows with the c.m. energy of the colliding electrons, $\sigma^{Ztc}$ falls as $\sqrt s$ grows. 
Therefore, a search for the $Ztc$ signature will be most effective at lower energies. 

\begin{figure}
\psfig{figure=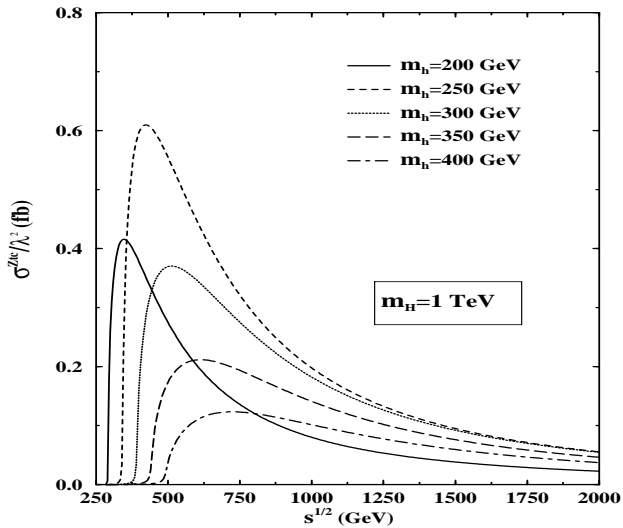,height=2.0in,width=4.0in}
\caption{The cross-section $\sigma (e^+e^- \to t \bar c Z + \bar t c Z)$ in units of $\lambda^2$ as a function of $m_h$ for center of mass energies of 0.5,1,1.5 and 2 TeV. 
\label{fig3}}
\end{figure}

\begin{figure}
\psfig{figure=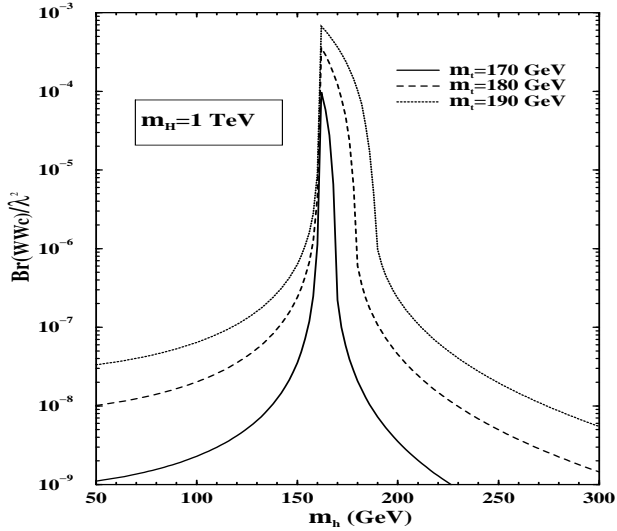,height=2.0in,width=4.0in}
\caption{The scaled branching ratio, $Br(t\to W^+W^-c)/\lambda^2$
as a function of $m_h$ for various values of $m_t$. 
\label{fig4}}
\end{figure}

\section{The Rare Top Decays $t \to W^+W^- c$, $t \to ZZc$}

Last we discuss the two rare decays $t \to W^+W^-c$
and $t \to ZZc$. The latter is, of course, possible only if $m_t > 2m_Z +m_c$. Within the SM these decay channels
are vanishingly small.
For the first one, the tree-level decay, ${\rm Br}(t \to W^+W^-c)
\approx 10^{-13}-10^{-12}$ due
to GIM suppression.\cite{jenkins} For the second decay the branching
ratio is even smaller since it occurs only at one loop. 

The situation is completely different in Model~III where both
decays occur at the tree-level through the FC Higgs exchange of
Fig.~1b. These decays are related to the fusion reactions
($WW$, $ZZ\to \bar tc$) by crossing symmetry.
Thus in terms of the hard cross-section given in Eq.~\ref{vvhtc}: 
\begin{equation}
\Gamma_V \equiv \Gamma(t \to V V c)= { m_t^3 \over 32 N_c \pi^2}
\int_{4\qwe_V^2}^{(1- \qwe_c )^2} \!\!\!\!\!\! dz ~
z (z - 4\qwe_V^2)  \!\!\! \!\!\! \sum_{h_{V^1},h_{V^2}} 
\left. {\hat {\sigma}}_V \right|_{\hat s = m_t^2 z}
\end{equation}
\noindent The scaled-branching-ratio (SBR) $ {\rm Br}(t \to
W^+W^-c)/ \lambda ^2$
is given in Fig.~4; it is largest for $ 2m_W \lsim m_h \lsim m_t$ and drops rapidly in the regions $m_h<2m_W$ or $m_h>200$ GeV. 
For a wide range of $m_h$, the SBR is
many orders of magnitude bigger than the SM. 
Indeed for optimal
values of $m_h$, lying in the very narrow window, $2m_W \lsim m_h \lsim m_t$, the SBR$\sim 10^{-4}$. It is typically a few times $10^{-7}$ for $m_h \gsim m_t$ and can
reach $ \sim 10^{-6}$ in the $m_h \lsim 2m_W$ region.  
Concerning $t \to ZZc$, the branching ratio is typically $\sim 10^{-5}$
for $(2m_Z +m_c) < m_t < 200$ GeV if again $m_h$ lies in a very narrow window, $2m_Z<m_h<m_t$.
Note that, in contrast to the SM, within Model~III, Br($h \to
WW$)$ \sim 1$ for ${\tilde {\alpha}}=\pi/4$; and, even for $ m_h > 2 m_t
$, Br($ h \to WW $)$ \sim 0.7 \gg $Br($ h \to t \bar t$).\footnote{Note that Hou's analytical results
 in Ref.~3 correspond to the choice $\tilde\alpha=0$ in our notation. In this
special case Higgs decays to $WW$, $ZZ$ are suppressed at tree level even when
$m_h>2m_W$. In contrast, for illustrative purposes, we are using
$\tilde\alpha=\pi/4$ in which case $h\to WW$ becomes the dominant
decay.}
Both decays are thus very sensitive to $ m_t $: for $170 ~{\rm GeV}<m_t<200 ~{\rm
GeV}$, a $\sim 15$ GeV shift in $ m_t $ can generate an order of magnitude
change in the Br in the region $2m_V<m_h<m_t$.

\section{Summary and Conclusions}

To summarize, we have emphasized the importance 
of searching for the FC reactions, $e^+e^- \to t \bar c \nu_e {\bar {\nu}}_e$ 
, $e^+e^- \to t \bar c e^+ e^-$ and $e^+e^- \to t \bar c Z$, in a high energy $e^+e^-$ 
collider. These reactions are sensitive indicators of physics 
beyond the SM with new FC couplings of the top quark. As an 
illustrative example we have considered the consequences of extending 
the scalar sector of the SM with a second scalar doublet such that new 
FC couplings occur at the tree-level. For the most promising ones, the 
$t \bar c \nu_e {\bar {\nu}}_e$ and $t \bar c e^+ e^-$ final states, we found that within a large 
portion of the free parameter space of the FC 2HDM, these new FC 
couplings may give rise to appreciable production rates which can unambiguously indicate the existence of new physics. 
In particular, in a NLC running at energies of 
$\sqrt s \gsim 1$ TeV and an integrated luminosity of the order 
of ${\cal L} \gsim 10^2$ [fb]$^{-1}$, Model~III (with $\lambda=1$) predicts hundreds and 
up to thousands of $t \bar c \nu_e {\bar {\nu}}_e$ events and several tens  
to hundreds $t \bar c e^+ e^-$ events.


\begin{thebibliography}{99}

\bibitem{hepph9703221} S. Bar-Shalom, G. Eilam, A. Soni and J. Wudka, 
hep-ph-9703221, Report no.: UCRHEP-T185 (UC Riverside, 1997).

\bibitem{model3} S. Glashow and S. Weinberg, \Journal{\PRD}{15}{1958}{1977}; M. Luke and M.J. Savage, \Journal{\PLB}{307}{387}{1993}; D. Atwood, L. Reina and A. Soni, hep-ph-9609279.

\bibitem{model3phen} M.J. Savage, \Journal{\PLB}{266}{135}{1991};
W.S. Hou, \Journal{\PLB}{296}{179}{1992}; L.J. Hall and S.
Weinberg, \Journal{\PRD}{48}{R979}{1993}; D. Atwood, L. Reina and A. Soni, 
\Journal{\PRL}{75}{3800}{1995}, \Journal{\PRD}{53}{1196}{1996}; 
W.-S. Hou and G.-L. Lin, \Journal{\PLB}{379}{261}{1996}.

\bibitem{csa} T.P. Cheng and M. Sher, \Journal{\PRD}{35}{3484}{1987};
M. Sher and Y. Yuan, \Journal{\PRD}{44}{1461}{1991}.

\bibitem{jenkins} E. Jenkins, hep-ph-9612211, Report-no.: 
UCSD/PTH 96-28 (UC San Diago, 1996); D. Atwood, Jefferson Laboratory and M. Sher, College 
of William and Mary, private communication (1997).

\end{thebibliography}
\end{document}